\documentclass[%
reprint,
superscriptaddress,
showpacs,preprintnumbers,
 amsmath,amssymb,
 aps,
]{revtex4-1}
\usepackage{graphicx}
\usepackage{dcolumn}
\usepackage{bm}
\usepackage{amsmath}%
\usepackage{amsfonts}%
\usepackage{amssymb}%
\usepackage{subfigure}%

\begin{document}

\preprint{submitted to PRL}

\title{Enhanced Proton Acceleration by an Ultrashort Laser Interaction with Structured Dynamic Plasma Targets}

\author{A. Zigler}
\author{S. Eisenman}
\author{M. Botton}
\email[Corresponding author: ]{bdmoti@phys.huji.ac.il}
\author{E.Nahum}
\author{E. Schleifer}
\author{ A. Baspaly}
\author{Y.Pomerantz}
\affiliation{%
 Racah Institute of Physics, The Hebrew University of Jerusalem
 Jerusalem, ISRAEL
}%


\author{F. Abicht}
\author{J. Branzel}
\author{G.Priebe}
\author{S.Steinke}
\author{A. Andreev}
\author{M. Schnuerer}
\author{W.Sandner}
\affiliation{
 Max Born Institute, Berlin ,Germany
}%
\author{D. Gordon}
\author{P. Sprangle}
\affiliation{%
 Plasma Physics Division, Naval Research Lab, Washington, D.C. 20375, USA
}%
\author{K.W.D.Ledingham}
\affiliation{
University of Strathclyde,Glasgow G4 0NG, Scotland, UK
}

\date{\today}

\begin{abstract}
We experimentally demonstrate a notably enhanced acceleration of protons to high energy by relatively modest ultrashort laser pulses and structured dynamical plasma targets. Realized by special deposition of snow targets on sapphire substrates and using carefully planned pre-pulses, high proton yield emitted in a narrow solid angle with energy above 21MeV were detected from a 5TW laser. Our simulations predict that using the proposed scheme protons can be accelerated to energies above 150MeV by 100TW laser systems. 
\end{abstract}

\maketitle


	Proton acceleration by the interaction of an ultra high intensity 
laser beam with matter has several wide prospective applications including cancer treatment,
astrophysics in the lab, nuclear physics, and material sciences (see [1,2] for review). 
Over the years several promising acceleration schemes were proposed and demonstrated, 
such as Target Normal Sheath Acceleration (TNSA) [3-5], 
Radiation Pressure Acceleration (RPA) [6,7], Break Out Afterburner (BOA) [8], 
and collisionless shock acceleration [9]. Variations of these schemes including mass-limited 
targets [10] nano-structure targets [11-13] aimed at increasing the efficiency of the interaction 
were also considered. Some of these schemes require exceedingly a high intensity of the laser 
beam in order to initiate the process, and most of them require a laser energy exceeding 1PW 
level (on target) in order to accelerate the protons to energies about 150MeV required by 
medical applications. Furthermore, all these schemes pose the same requirement of the laser 
system to have as low as possible energy stored in the pre-pulse. The main reason is that these
accelerating schemes are optimized for an interaction between the main pulse and a cold 
solid-density target, and are strongly degraded as the main laser pulse interacts with a 
pre-heated (or even ionized) target. It was recently shown that pre-formed plasma may be beneficial to the acceleration process [14,15]. Nevertheless, the majority of the experiments still 
aim at cold targets with as low as possible pre-pulse. Accordingly, achieving such a low energy
content in the pre-pulse requires a contrast ratio of the order of 10$^{-11}$ for a 100TW 
laser and even higher for more energetic systems, which is a real experimental challenge. 

	In this letter we report for the first time on acceleration of proton bunches to energies 
above 21MeV by  a 5TW ultrashort (50fs) laser pulse. 
We introduce an alternative approach to the laser-based 
proton acceleration quest by using a moderate power ($<$10TW) laser system, and carefully 
produced microstructured snow targets [16-17]. Our experimental results show that the energy 
of the accelerated protons scales with the power of the laser according to 
the $E_p \propto {P_L}^{1/2}$ rule which is obtained
here for significantly lower laser powers than the traditional schemes. Numerical 2D PIC code
simulations of the interaction process reproduce the experimentally obtained scaling law and predict
the possibility of accelerating protons to 150MeV with a laser power of about 100TW. This notably
increased proton energy is attributed to a combination of three mechanisms. First is the localised
enhancement of the laser field intensity near the tip of the microstructured whisker. This causes an
increased electronic charge repulsion out of the whisker. Second is a mass-limited like phenomena,
namely the absence of high density cold electron cloud in the vicinity of the whisker which can
compensate for the expelled electrons. The heated electrons remain in the vicinity of the positively
charged whisker, producing strong accelerating electrostatic fields and pulling the protons out. Third
is the Coulomb explosion of the positively charged whisker, adding a longtime acceleration to the
protons. As our innovative microstructured snow scheme requires the interaction of the laser with a
structured dynamical plasma target, it also relaxes the requirements of high contrast ratio of the laser
system, and facilitates the production of the target.		
	
	Two separate laser systems were used to obtain the results reported here. One is the laser at the
High-Intensity Laser Lab at the Hebrew University of Jerusalem (HUJI) and the other is the laser
system at the Max Born Institute at Berlin (MBI). The HUJI system is a 2TW, 50fs pulse length laser
operating at a central wavelength of 0.8 $\mu$. The spot size (FWHM) as small as 20 $\mu m^2$. 
The contrast ratio of the HUJI system is about 1000 dictated by the leakage of the  regenerative amplifier, thus the
pre-pulse is shaped similarly to the main pulse and arrives on the target approximately 10nsec before
the main pulse. The MBI system is a 30TW, 70fs 0.8 $\mu$ laser. The beam is focused to a spot size of
5$\mu$  on target. The contrast ratio of the MBI system is about 10$^{-8}$, and the pre-pulse is a pedestal
like shape. In order to obtain the same pre-plasma conditions in both institutes, the main beam at the
MBI system is split and a pre-pulse is artificially formed. The experiments at the MBI site were carried
out using a 0.03mJ prepulse 6ns ahead of the main pulse.	

	A schematic diagram of the experimental setup in both institutes is shown in Fig. 1. The laser
beam strikes the target at an angle 60$^\circ$ to the normal. The target is a liquid nitrogen cooled sapphire
substrate on which water vapor is deposited. The size and density of the snow pillars as well as the
surface roughness are determined by various parameters such as pressure, flow and temperature. 
\begin{figure}
  \includegraphics[width=0.48\textwidth]{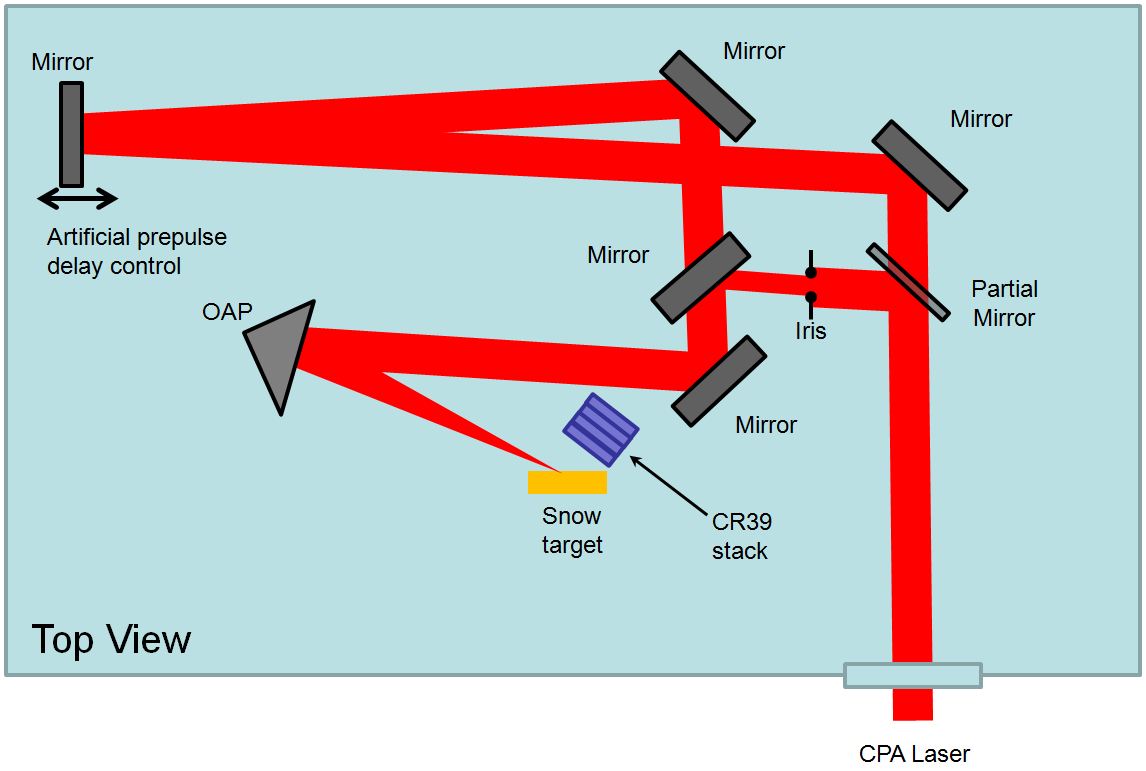}
\caption{\label{fig:wide}The experimental setup}
\end{figure}

Fig. 2 shows a typical snow targets as imaged by a scanning electron microscope. 
The snow surface can be characterised as a highly structured surface with essentially three
roughness scales: a) pillars of about 100$\mu$, b) spikes of about 10$\mu$ on top of them 
and c) whiskers of about 1$\mu$ on the spikes. 
The distribution of the pillars and spikes is controlled by seeding and fixing
the flow of the vapor.  The red circle in Fig 2 schematically represents the laser spot (pre-pulse and
main pulse) for a typical shot. Based on our measurements of the laser spot characteristics, we
estimate that the laser spot interacts with few (typically less than 5) whiskers. 

\begin{figure}
\centerline{
 \includegraphics[width=0.24\textwidth]{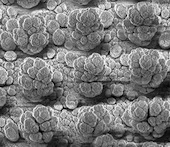}
 \includegraphics[width=0.24\textwidth]{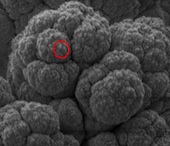}
}
\caption{\label{fig:wide}Scanning electron microscope images of the snow targets. Red circle represents the laser spot. }
\end{figure}

The pre pulse meets the whiskers and vaporizes part of them. The time delay between the pre and
main pulse is about 10ns, during which the plasma can freely expand and a highly non-uniform
plasma cloud is formed. By the time of its arrival, the main pulse meets this highly structured dynamic
plasma cloud and interacts with it to produce the accelerated protons. 	The main diagnostics
consists of stacks of CR39 film, each 25 mm diameter. The thickness of the CR39 is 1.5mm. In front of
the first CR39 film we positioned a 13$\mu$ Aluminum foil which prevents the plasma plume emitted in the
process to damage the films. The detector films are positioned at a distance of about 2.5cm from the
target at an angle of 450. The setup enables sufficient energy and spatial resolution. The CR39 plates
are replaced every several laser shots (typically 3-5) and the exposed films are etched using the
conventional methods. Analysis of the processed CR39 consisted on automated counting of the pits
and comparing the resulted number to unexposed areas on the same film and on a reference
(processed) film. Exposed areas in all the films showed clear signal far above the noise level, which is
unquestionably the result of accelerated protons.  Fig. 3 shows a 100$\mu$ x 100$\mu$ area of three 
processed CR39 films. The first plate was uniformly covered by proton tracks. Considering the
thickness of the foil that covered the stack it is concluded that the tracks are produced by protons of
energy above 1MeV.  The second CR39 detector was covered by a lower density but still uniform
distribution of  tracks. These are attributed to protons of energy above 13MeV. The tracks marked on
the third plate are the result of protons with energy above 20MeV. 
Here the density is not uniform and bunches can be seen. 

\begin{figure}
\centerline{
 \includegraphics[width=0.16\textwidth]{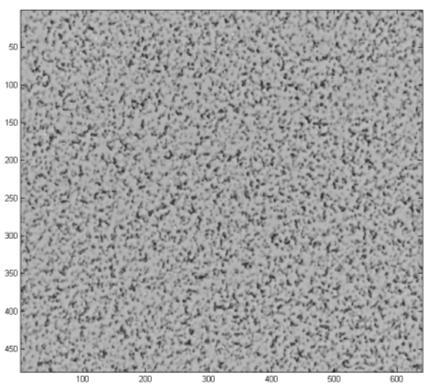}
 \includegraphics[width=0.16\textwidth]{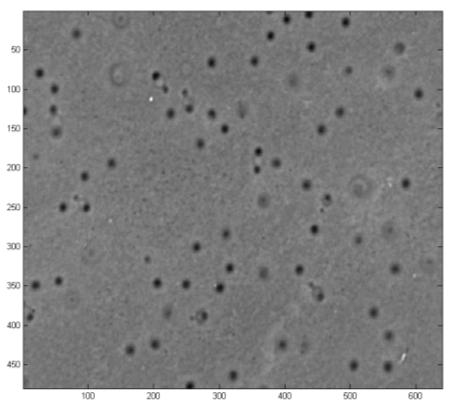}
 \includegraphics[width=0.16\textwidth]{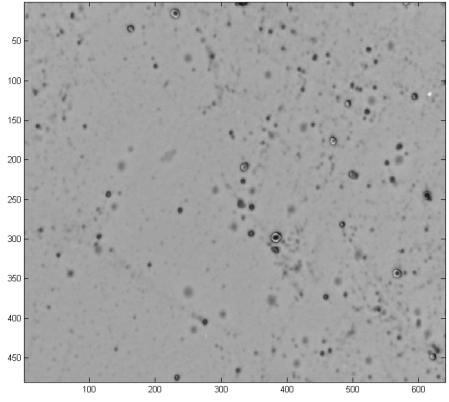}
}
\caption{\label{fig:wide}Processed CR39 showing marks of protons with energy $>$1MeV. (left), 
energy $>$13MeV. (middle) energy $>$20MeV. (right)}
\end{figure}

We have repeated the experiment with various laser power levels to produce the energy scaling of the
most energetic protons accelerated of the snow targets (Fig. 4).   The fitted scaling from this data set of
this novel acceleration scheme is $E_p \propto {P_L}^{0.58}$, and is obtained with lower laser power levels. Notice that unlike
the traditional TNSA scheme there is no clear distinction of back and front surface as the laser does
not cross the target but interacts with its micro structured surface and the plasma cloud generated
there by the pre pulse. At this point we note that when the pre-pulse was not present and the main
pulse interacted directly with the snow surface, the resulted protons were accelerated to a much lower
energy level, down to the traditional scaling of the TNSA scheme. A count of the total number of tracks
on the CR39 detector plates is shown in the inset of Figure 4. The measured energy spectrum is not a
monotonically decreasing function of the energy (within the available resolution) and a clear peak is
evident at energies around 20MeV. We also found that the most energetic protons were found on the
CR39 detectors at a location which is close to the position of a reflected laser field for a planar target.
\begin{figure}
 \includegraphics[width=0.49\textwidth]{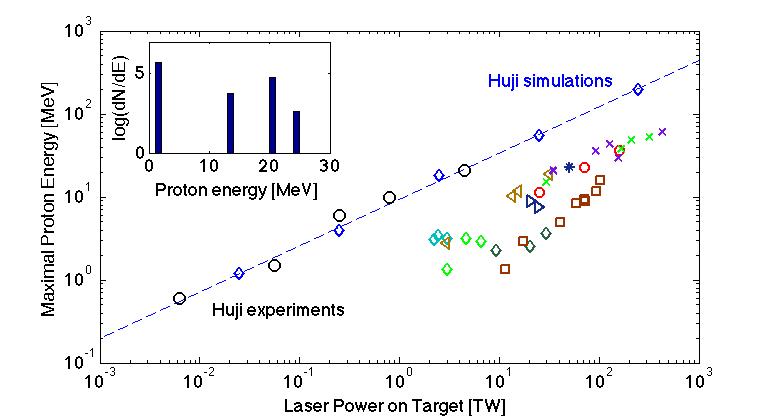}
 \caption{\label{fig:wide}Energy scaling of the accelerated protons as a function of the laser 
power on target. Various points describe data obtained by many contributors (See ref [18] for details)}
\end{figure}

	The TURBOWAVE particle in cell (PIC) [19] simulations of the unique acceleration scheme
presented here are based on the one dimensional model presented in our previous work [17] which
demonstrated the significant enhancement by the local plasma density near the whisker tip.
Considering the size scales of the target and the laser spot size we focus our study on the interaction
of the laser with a single snow whisker. Following the pre-pulse illumination, the whisker is partially
vaporized and ionized hence a non uniform plasma cloud (protons, triply-ionized oxygen and
electrons) is formed.We model this highly structured dynamic plasma as an ellipsoid.  We estimate
that the high ($\thicksim100n_{cr}$, where $ n_{cr}=1.1\cdot10^{21}/ \lambda^2 $ is the 
traditional critical density) density portion of the
whisker is an ellipsoid of the order of $0.1-0.2\mu$ width (minor axis) and $1-2\mu$ l
ength (major axis).
Estimates of the density gradients of the plasma cloud (supported by a 1D hydrocode Hyades [20])
setthe critical density contour to be an ellipsoid of the order of $1-2\mu$ width (minor axis) 
and $10\mu$ length (major axis). The linearly polarized laser has a spot size of $5\mu$ 
and is taken to hit the plasma cloud at an angle of $ 45^{\circ}$ with respect to the major axis. 
The laser pulse is 88fs total duration with a linear rise
time and fall time of 32fs each. Further simplification is achieved by eliminating one dimension and
setting the whisker to be independent of the y coordinate (hence it becomes a wedge), keeping the
polarization of the laser in the x-z plane. The 2D model is expected to reproduce the energy scaling of
the accelerated protons but not the spatial distribution which will be strongly dependent on the angle
between the polarized field and the major axis of the plasma ellipsoid as well as other neighboring
whisker which affect the propagation of the laser field.The cell size is $0.05\mu \times 0.05\mu$, 
and the simulation region is $200\mu \times 200\mu$. 

\begin{figure}
\centerline{
 \includegraphics[width=0.16\textwidth]{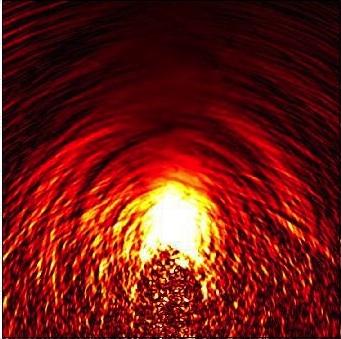}
 \includegraphics[width=0.16\textwidth]{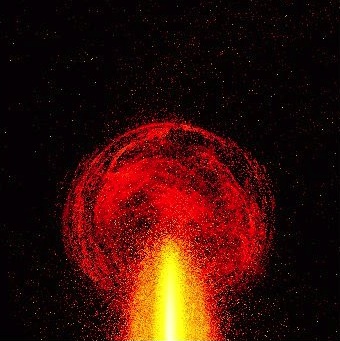}
 \includegraphics[width=0.16\textwidth]{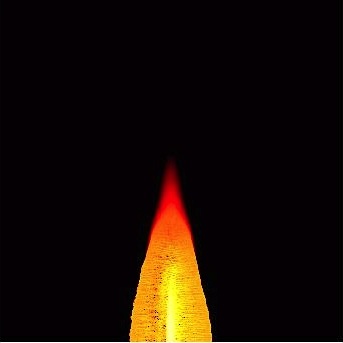}
}
\centerline{
 \includegraphics[width=0.16\textwidth]{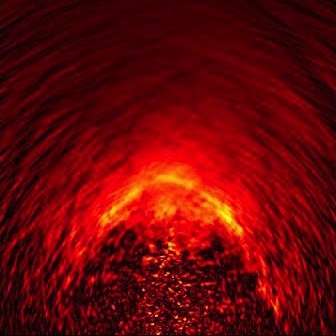}
 \includegraphics[width=0.16\textwidth]{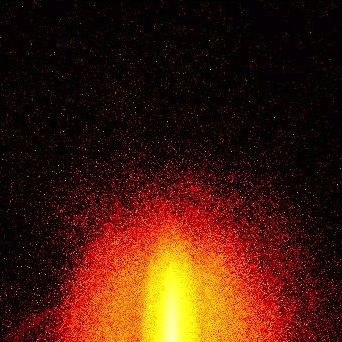}
 \includegraphics[width=0.16\textwidth]{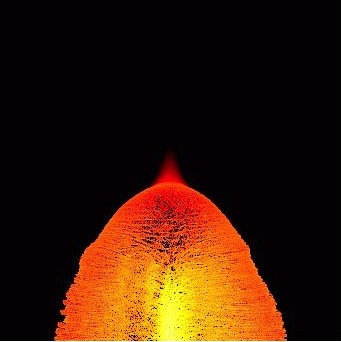}
}
\caption{\label{fig:wide}Spatial distribution of protons (right), electron (middle), and electric field
intensity along the major axis (left) at 220fs (top row) and 440fs (bottom row). 
All frames are $40\mu \times 40\mu$ .Colormap scale for density are logarithmic 
normalized to critical density, and for field intensity are in normalized 
units $(a_0=eE/mc\omega)$}
\end{figure}

	The interaction and acceleration process is roughly separated into three phases. The first phase,
which is approximately the laser pulse duration, the laser drives the electrons out of the plasma
ellipsoid. These electrons are accelerated by the laser's ponderomotive potential and part of them
escape out of the plasma cloud leaving the heavy highly charged oxygen atoms and the protons
behind. The curved equal density surfaces affect the ponderomotive potential force. Unlike the TNSA
target, there is no clear distinction between back and front surfaces as they are mixed over the
whisker tip. The second phase is a short time after the passage of the laser pulse, approximately 2-3
pulse duration. The top row of Fig. 5 shows the density of the protons (right) electrons (middle) and
and the electric field along the major axis (left) at 220fs. Here we see the electron accumulation near
the tip of the whisker. The protons starts to react to the accelerating field. The accumulation of the
electrons by the tip produces a strong electrostatic field. The third and last phase is long time
acceleration, around 3-6 laser pulse duration. The electron accumulation near the whisker tip is
reduced but charge separation between electrons and ions is maintained. Fig 5 bottom row shows the
density of proton, electron, and electric field along the major axis at 440fs. The charge separation
between the hot electrons and the protons generates an accelerating field while the un-neutralized
heavier oxygen ions remain behind and add a pushing field. The highest energy of the protons for this
simulation was found to be 15MeV. The scaling law of figure 4 was obtained by repeating the
simulations with changed laser power. 

	In conclusion, we have experimentally demonstrated acceleration of protons to energy up to
25MeV by a modest power  $(\thicksim 5TW)$ laser. The acceleration scheme is based on 
a highly structured dynamic plasma target which is produced by a pre-pulse illumination of a 
snow target. The energy
scaling resembles the power law of planar targets but obtained with a much lower laser energy. The
proton yield is of the order of $10^6$ protons per shot with an angular distribution of the high energy
component of protons is $(\thicksim0.1 radians)$. PIC simulations reproduce the 
energy scaling and predict that
by using 100TW laser, protons can be accelerated to 150MeV level. This is a considerable
improvement over alternative acceleration schemes which requires a higher energy laser system with
an extremely high contrast ratio.

\nocite{*}


\end{document}